\documentclass[acmsmall,screen,authorversion,nonacm]{acmart}

\usepackage{multirow}

\AtBeginDocument{%
  \providecommand\BibTeX{{%
    \normalfont B\kern-0.5em{\scshape i\kern-0.25em b}\kern-0.8em\TeX}}}

\begin{document}

\title{A Modular End-to-End Framework for Secure Firmware Updates on Embedded Systems}

\author{Solon Falas}
\affiliation{%
  \institution{Dept. of Electrical and Computer Engineering and KIOS Research and Innovation Centre of Excellence, University of Cyprus, Nicosia\authornotemark[1]}
  \country{Cyprus}
}
\email{falas.solon@ucy.ac.cy}

\author{Charalambos Konstantinou}
\affiliation{%
  \institution{KAUST, Thuwal\authornotemark[2]}
  \country{Saudi Arabia}
}
\email{charalambos.konstantinou@kaust.edu.sa}

\author{Maria K. Michael}
\affiliation{%
  \institution{Dept. of Electrical and Computer Engineering and KIOS Research and Innovation Centre of Excellence, University of Cyprus, Nicosia\authornotemark[1]}
  \country{Cyprus}
}
\email{mmichael@ucy.ac.cy}

\begin{abstract}
Firmware refers to device read-only resident code which includes microcode and macro-instruction -level routines. For Internet-of-Things (IoT) devices without an operating system, firmware includes all the necessary instructions on how such embedded systems operate and communicate. Thus, firmware updates are an essential part of device functionality. They provide the ability to patch vulnerabilities, address operational issues, and improve device reliability and performance during the lifetime of the system. This process, however, is often exploited by attackers in order to inject malicious firmware code into the embedded device. In this paper, we present a framework for secure firmware updates on embedded systems. The approach is based on hardware primitives and cryptographic modules, and it can be deployed in environments where communication channels might be insecure. The implementation of the framework is flexible as it can be adapted in regards to the IoT device's available hardware resources and constraints. Our security analysis shows that our framework is resilient to a variety of attack vectors. The experimental setup demonstrates the feasibility of the approach. By implementing a variety of test cases on FPGA, we demonstrate the adaptability and performance of the framework. Experiments indicate that the update procedure for a $1183kB$ firmware image could be achieved, in a secure manner, under $1.73$ seconds.
\end{abstract}


\maketitle

\section{Introduction}
With the advancement of Internet-of-Things (IoT) technology, Embedded Devices (EDs) have increasingly permeated our daily lives. Such devices allow physical objects to become integrable with information. As a result, EDs are increasingly interdependent and in many cases important to our safety. For example, critical domains such as Industrial Control Systems (ICS) are being integrated into Industrial IoT (IIoT) embedded systems in order to augment traditional control systems with wireless sensing services and provide better automation through multiple sensors and measurement points. This massive deployment of EDs in mission-critical environments introduces security challenges. 
Such devices are highly constrained in terms of performance and resources, hence it is often infeasible to employ traditional security techniques as those used in general-purpose computing systems. The number of incidents causing software failures, data breaches, and often physical damage is increasing. For instance, malicious adversaries can compromise high-wattage IoT devices to disrupt the power grid's normal operation by manipulating the total load demand  \cite{soltan2018blackiot}.

\subsection{Motivation}
Firmware is commonly referred to as ``the operating system of an ED'' \cite{basnight2013firmware}. It is the dedicated software residing in read-only memory, a layer between software and bare-metal hardware. It provides low-level control of the device. Firmware has to be routinely updated in order to fix bugs, address performance issues, and enhance the functionality of a device.
Recent surveys quantify that there are no significant security gains in firmware security in recent years \cite{citl2019,thompsonzatko2018}. Firmware continues to be an enticing entry point for attackers. IoT devices are often low-cost and produced in high volume. Moreover, their development time is minimal. Therefore, security is typically  considered of low priority in the production pipeline.
In addition, the process of securely updating the firmware of EDs while minimizing system downtime is still an issue for many system administrators and manufacturers \cite{houstonchr}. 
Mass production of devices manufactured in the IoT domain may impair cybersecurity efforts. IoT firmware security is typically not being addressed to the same level as that of general-purpose computers and BIOS security \cite{mocker2019}. 
Lack of proper security mechanisms during the firmware delivery and update process may allow malicious adversaries to gain full control of a device. Consequently, swarms of compromised IoT devices of different functionality and purpose can launch coordinated Denial-of-Service (DoS) attacks, e.g., Mirai botnet \cite{sealstara}. Security can no longer be an afterthought.

Despite efforts of software-based security approaches, malicious threats are growing. These threats exploit bugs in the operating systems, user applications, and the software itself.
In order to address these challenges, our proposed framework ensures firmware data integrity and confidentiality by leveraging: (1) hardware-based cryptographic primitives, and (2) hardware-intrinsic characteristics, to perform authentication procedures.
Effective security solutions can utilize the underlying hardware, avoiding storage of sensitive information in non-volatile memories.
Chip-specific characteristics act as ``digital fingerprints'' that pair each firmware image to the correct IoT device recipient while the hardware crypto-cores handle the delivery of the encrypted firmware. 
Our technique ensures that IoT devices, often exposed to external threats, can have their firmware code securely updated remotely. The cryptographic elements being used in this procedure include hash functions and encryption algorithms. The means of digital fingerprinting are demonstrated via Physical Unclonable Functions (PUFs) 
\cite{joshi2017everything}.
The firmware updates for IoT devices rely on hardware as a root of trust, thus avoiding reliance on pre-stored software routines.

\subsection{Contributions}
The proposed framework is motivated towards low-end EDs, especially IoT devices typically deployed in industrial settings and critical infrastructures. These devices are often installed in remote environments where physical access might be possible. In this framework, we describe a secure protocol that integrators (e.g., system designers, IoT infrastructure owners, etc.) can utilize for remote firmware updates.
We propose an end-to-end secure and modular framework incorporating two-way authentication handshakes, strong confidentiality guarantees, and protection mechanisms against a variety of possible attacks. Our proposed methodology leverages hardware primitives to deliver firmware updates and provides significant advantages over existing PUF-based techniques used primarily for authentication purposes. In particular:

\begin{itemize}
    \item We propose an \textit{end-to-end} framework for secure firmware updates on embedded systems. The framework does not require storing secret information in the device's non-volatile memory. It focuses on data confidentiality, integrity, and authenticity while providing mechanisms that ensure the device's availability. The framework includes a fast and secure firmware update delivery protocol that leverages hardware-implemented cryptographic primitives to provide high levels of security against a variety of attacks. Furthermore, it can enroll new devices in hostile environments without prior set up in a secure environment. The framework is also flexible in terms of cryptographic component choice as such components can be interchanged to meet different constraints and requirements. 
    \item The effects of manufacturing variability and the concept of Public PUFs (PPUFs) are utilized for authentication purposes. The PPUFs' unique properties allow for device enrollment without prior setup or onboard secrets, a unique feature presented in this work. In addition, PPUFs are leveraged for making the firmware update packages chip-specific. Therefore, compromising a firmware update for a specific device will not impact the security of same-model devices.
    \item The security of the approach is evaluated using \textit{STRIDE}, a well-known systematic threat modeling approach, demonstrating strong security guarantees against a variety of attacks. The detailed design process describes the goals and \textit{Secure Design Requirements} (SDRs) set for maintaining high performance and security while requiring minimal hardware resources. STRIDE helps in effectively demonstrating how each attack is thwarted by SDRs and exhibits the protection provided by the features that our protocol employs. We also perform extensive security analysis on additional sophisticated attack vectors in order to formally prove the security robustness of our proposed framework.
    \item We develop a \textit{modular} framework which allows us to consider different configurations. Specifically, we examine three indicative configurations catering to the diverse device-space of IoT. These configurations, provide different trade-offs in terms of performance, hardware resources required, and security level. Well documented hardware-implemented security algorithms are employed in the design of these configurations. The extensive experimental setup evaluation takes into account these hardware configurations by using industrial off-the-shelf firmware binaries in order to evaluate the feasibility and applicability of the approach. The results show promising performance while keeping significantly high levels of security.
\end{itemize}

\begin{figure}[t]
    \centering
    \includegraphics[scale=0.20]{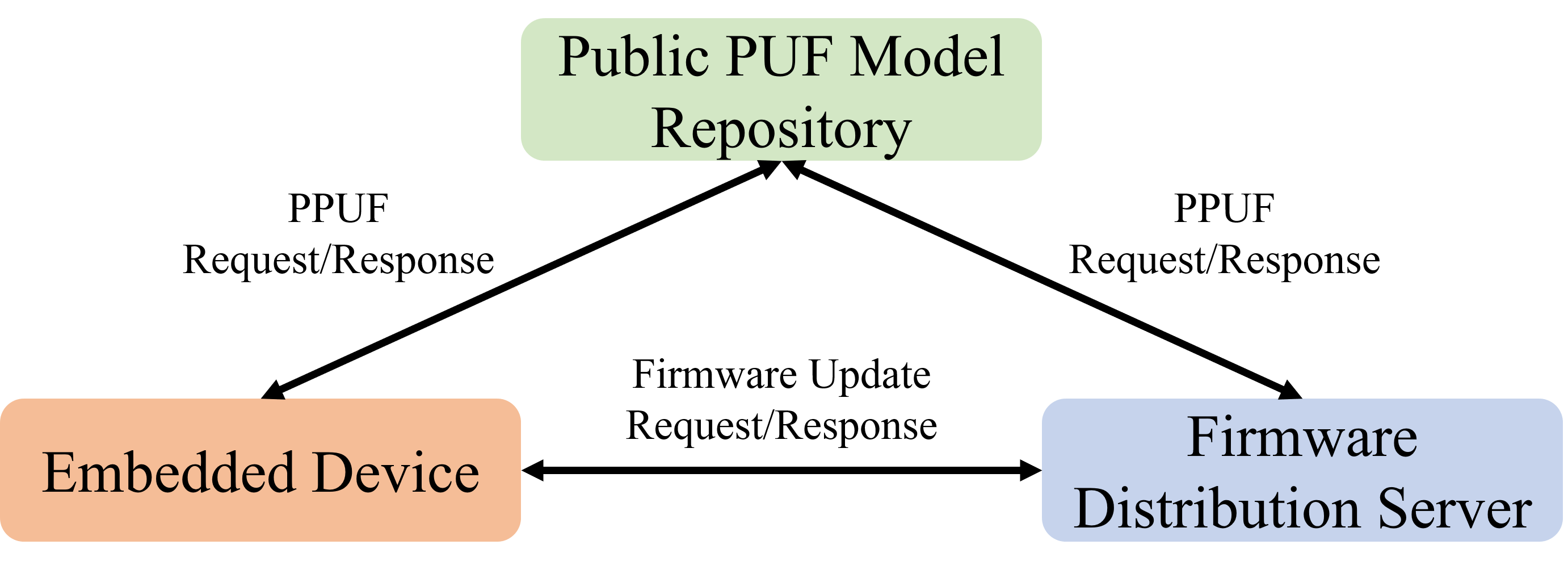}
    \caption{Overview of proposed concept: the PPUF-enabled IoT device requests a firmware update from the distribution server. The server replies with an encrypted firmware package. A public PUF model repository aids in key-creation on both sides and enables the secure transmission of the firmware image.}
    \label{fig:protocol summary}
\end{figure}

An overview of the proposed concept and high-level system architecture is presented in Fig. \ref{fig:protocol summary}.
The ED requesting the firmware update and the Firmware Distribution Server (FDS), containing the firmware updates, are two PPUF-enabled devices. Specifically, both units support the same hardware accelerators for hashing and encryption/decryption as well as the same-type of PPUFs.
The Public PUF Model Repository (PPMR)  contains non-sensitive and publicly accessible PPUF models for both devices. Using these models, the ED and FDS can challenge each other's PPUFs to perform mutual authentication, in the process of establishing a secure communication channel between them.
The proposed protocol enhances this procedure with data encryption and integrity mechanisms in order to realize a secure firmware update delivery from the FDS to the requesting ED.

\subsection{Paper Organization}
The rest of the paper is organized as follows. The security primitives considered in our framework are explained in Section \ref{s:background}. The proposed framework for securing firmware updates in EDs is presented in Section \ref{s:methodology}. The security of the framework against several attack methods is analyzed in Section \ref{s:security Analysis}. Our testbed and measurements are presented in Section \ref{s:experiments}. Results are compared to related work on PPUF-based firmware update security in Section \ref{s:related work}. Section \ref{s:conclusions} concludes the paper.

\section{Background}\label{s:background}

Our framework utilizes cryptographic routines implemented in hardware including encryption algorithms and cryptographic hash functions. It also uses the concept of PPUF, and its input-output behavior due to the random variations of the manufacturing process, as a hardware root-of-trust for authentication.

\begin{figure}[t]
    \centering
    \includegraphics[scale=0.074]{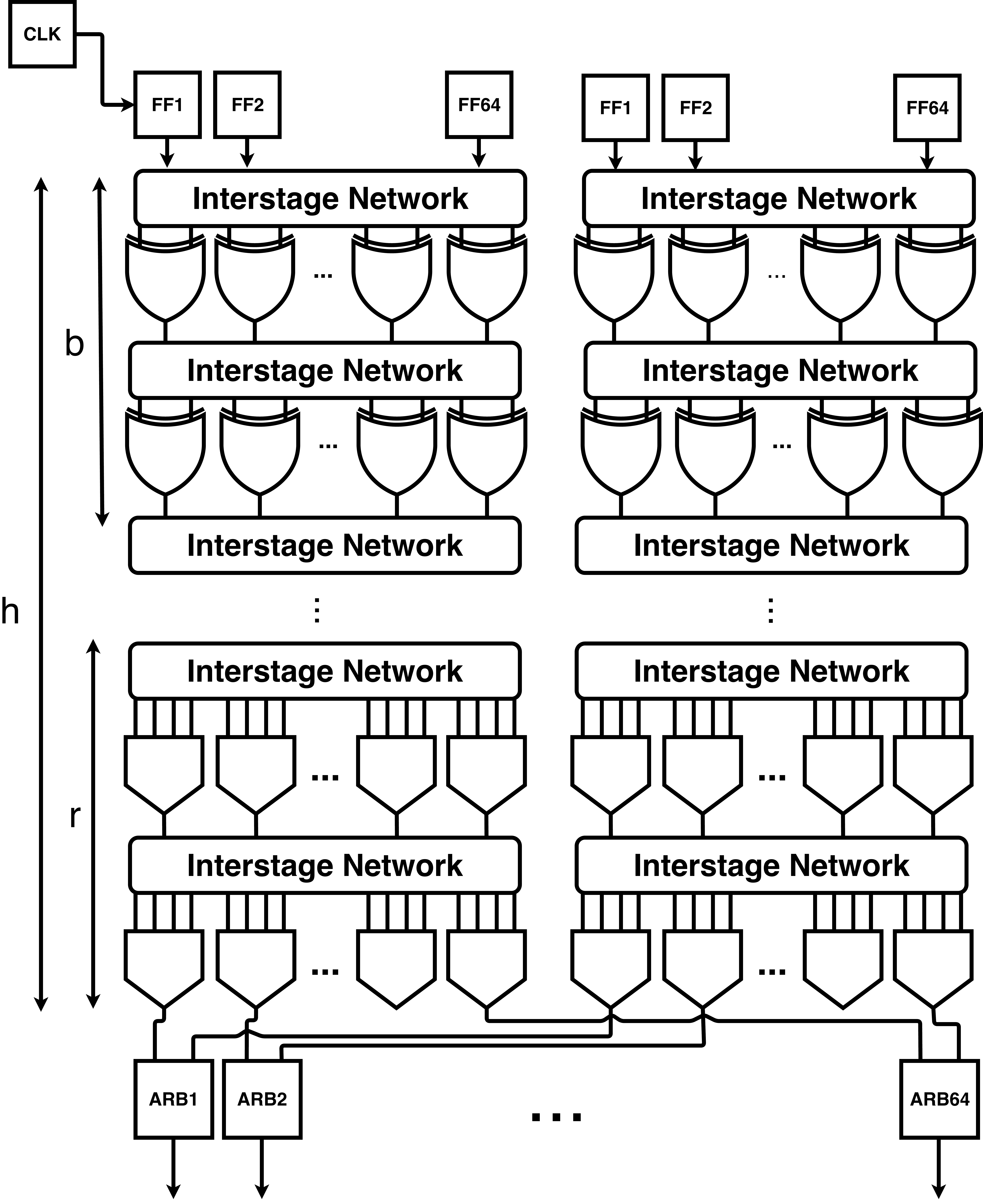}
    \caption{The differential Public PUF (dPPUF) is a series of booster and represser layers creating two identical structures (left part and right part). The two sides, however, are different in inherent delays (inertial, propagation, switching, etc.) due to manufacturing variability. The final arbiter layer captures the fastest propagating signals and creates the appropriate unique PUF response \protect \cite{potkonjak2011differential}.}
    \label{fig:dppuf}
    \vspace{-3mm}
\end{figure}

PPUFs are a type of PUFs that doesn't rely on the secrecy of their internal IC characteristics. They are designed to be fast-to-execute and slow-to-simulate systems \cite{potkonjak2014public}. In this work, we employ the differential PPUF (dPPUF) as the PPUF for our framework. dPPUFs do not require accurate timing and triggering mechanisms due to a layer of arbiters present as the last layer of gates before the dPPUF's output. The exact architecture of the dPPUF in our methodology can be seen in Fig. \ref{fig:dppuf}. A 256-bit version of the dPPUF is adopted from \cite{potkonjak2011differential}. Since PUFs exhibit inherent noise, their responses require some kind of error correction mechanisms to stabilize them. Error correction codes, special PUF designs, and other approaches have been proposed \cite{colombier2017key} to address this issue. In this work, without any loss of generality, we consider a Bose–Chaudhuri–Hocquenghem (BCH)-based code-offset fuzzy extractor as an effective PUF error correction mechanism \cite{dodis2008fuzzy}.

The gate-level characterization\footnote{Gate-level characterization is the process of characterizing each gate of an IC in terms of its physical properties using lasers, micro-probing, and simulations \cite{koushanfar2008post}. Typical characteristics measured include gate width and length, and properties such as leakage power and switching power.} of the dPPUF is performed by the manufacturer. The resulting IC characteristics are used to form a software model, which is stored in a public repository. This software model acts as the ``public part'' of our update protocol, enabling faster key-sharing between the communicating parties while rendering key recreation infeasible due to the extreme simulation requirements.
Since the public software model repository and firmware update distribution servers are owned by the device manufacturer, we assume that their contents are legitimate and intact. The use of PPUF leverages this kind of setup to avoid the need for third-party authenticators. Beyond the hardware characterization phase of the PPUF, no enrollment procedure is required. Therefore, the firmware manufacturer can package and send an update to a device without having to pre-install sensitive data in the device's memory or pre-share keys in a safe environment before device deployment.

An input to the PPUF's model, i.e., a \emph{challenge}, produces an output bit-string, i.e., the \emph{response}. The challenge and its corresponding response are also referred to as a \textit{Challenge-Response Pair (CRP)}. This procedure can be completed relatively fast on dedicated PPUF hardware and the software model. However, the opposite procedure, deriving which challenge created a particular response, is a computationally intensive task. Thus, the actual PPUF circuit owner can utilize the hardware to complete this task in a much shorter time than any kind of model simulation. The gap created between the simulating party and the actual PPUF owner trying to derive which challenge creates a known response is known as the Execution-Simulation Gap (ESG). Due to ESG, the PPUF model can be stored publicly without any security implications. This timing difference in conjunction with the unclonable nature of PUFs serves as our authentication pillar. The ESG can be manipulated to give as much advantage to the PPUF owner over the simulating party as needed.
Key width, number of challenges that need to be calculated, and search space size are all factors that if increased, the gap will also increase. 

The ESG property gives the opportunity to openly share metadata that can only be utilized by a specified IoT device recipient. The recipient is the only one who can effectively utilize the metadata in order to recreate the necessary keys for establishing a secure communication channel. Timing constraints and thresholds, implemented throughout the protocol, assist each party to make sure that it is communicating with the correct machine and block unauthorized access.


\section{Proposed Framework}\label{s:methodology}

In this section, we provide the details of the proposed framework. The key concept is to construct a firmware update package containing the encrypted firmware image as well as metadata that the manufacturer provides. This information allows the ED to securely authenticate and decrypt the firmware image. Specifically, we leverage the properties of cryptographic primitives (including encryption modules, hash algorithms, and PPUF) in a series of steps to create a firmware update protocol that ensures secure transmission of firmware packages from providers to device.

\subsection{Secure Design Requirements} \label{ss:secure design requirements}
The protocol needs to ensure that the device-specific firmware image can reach the appropriate embedded system while being protected by malicious interference during transmission assuming an insecure communication channel. Therefore, the protocol design requires guarantees to provide firmware image integrity, while the update process should not leak any useful to the attacker information during data transmission. Also, in case a firmware package gets intercepted, it must prevent any unauthorized party from accessing its contents. In the event of a corrupted package, the device should also be able to detect it. 
Finally, if a firmware update fails to complete multiple times (either due to malfunction or malicious interference), a mechanism must detect it in order to prevent the issue from escalating to onboard firmware corruption or service disruption.

In order to achieve these objectives, the framework is designed using the Secure Design Requirements (\textbf{SDRs}) presented below:

\noindent \textbf{SDR1 -- Monotonic Sequence Numbers:} The timestamps in our firmware update protocol contain monotonically increasing sequence numbers, e.g., Coordinated Universal Time (UTC) timestamps such as UNIX epoch. 
The device utilizes such sequence numbers to accept only firmware packages that are accompanied by newer timestamps than the one already installed. These sequence numbers can be used in conjunction with the firmware image revision or version, to protect the device against firmware roll-back attacks.

\noindent \textbf{SDR2 -- Vendor-Device -type Identifiers:} Devices must be able to accept only firmware packages intended for them based on matching identifiers. The firmware package includes, besides the image, data that allows the device to ``identify'' the vendor, model, hardware revision, software revision, etc. A large enough PPUF allows for no-collision device-specific keys, hence making each firmware package chip-specific. This alleviates the need for product details such as serial numbers and MAC addresses, consequently minimizing the impact on firmware payload overhead.

\noindent \textbf{SDR3 -- Best-Before Timestamps:} Firmware must have an end-of-life date in order to avoid deprecated updates on devices that have been offline for some time. Updating a device to a firmware version that is not the latest, leaves the device prone to non-zero day vulnerabilities \cite{CVE-2017-5698}. Devices can check with the current UTC epoch timestamp whether the firmware is within its expected lifetime period or not, and reject firmware packages that have a best-before time smaller than the current time.

\noindent \textbf{SDR4 -- Signed Payload:} Firmware packages must be digitally signed to (1) verify the source of the incoming package, and (2) ensure data integrity. Signing a firmware package correctly means that the firmware provider sends proper identifying evidence to the receiver and enables corruption checks to the message.

\noindent \textbf{SDR5 -- Cryptographic Authenticity:} The encrypted firmware image, as well as all other parts of the package, must have a demonstrable way of proving its authenticity in order to protect the device from modification attacks. A cryptographic hash function digest utilized as a checksum ensures the data integrity of the firmware package.

\noindent \textbf{SDR6 -- Firmware Encryption:} The firmware image within the package must be encrypted. This step prevents attackers from getting access to its contents. Without access, the attackers cannot reverse engineer the firmware and uncover system functionality, identify possible exploitable backdoors or zero-day exploits, or simply modify the firmware to damage or control the ED. The proposed protocol incorporates encryption algorithms implemented in hardware that protect from unauthorized access to the firmware image.

\noindent \textbf{SDR7 -- Embedded Device Availability:} The embedded device must be available for legitimate users and detect malicious interference that might hamper its usability. An attacker may try to overload a device with firmware update requests in order to make it inaccessible. The device must be able to differentiate between legitimate and malicious firmware update activity. The protocol includes the security precautions required to protect the ED's operation from unauthorized disruption and attacker-enforced downtime.

\begin{figure*}[!t]
    \centering
    \includegraphics[width=\textwidth]{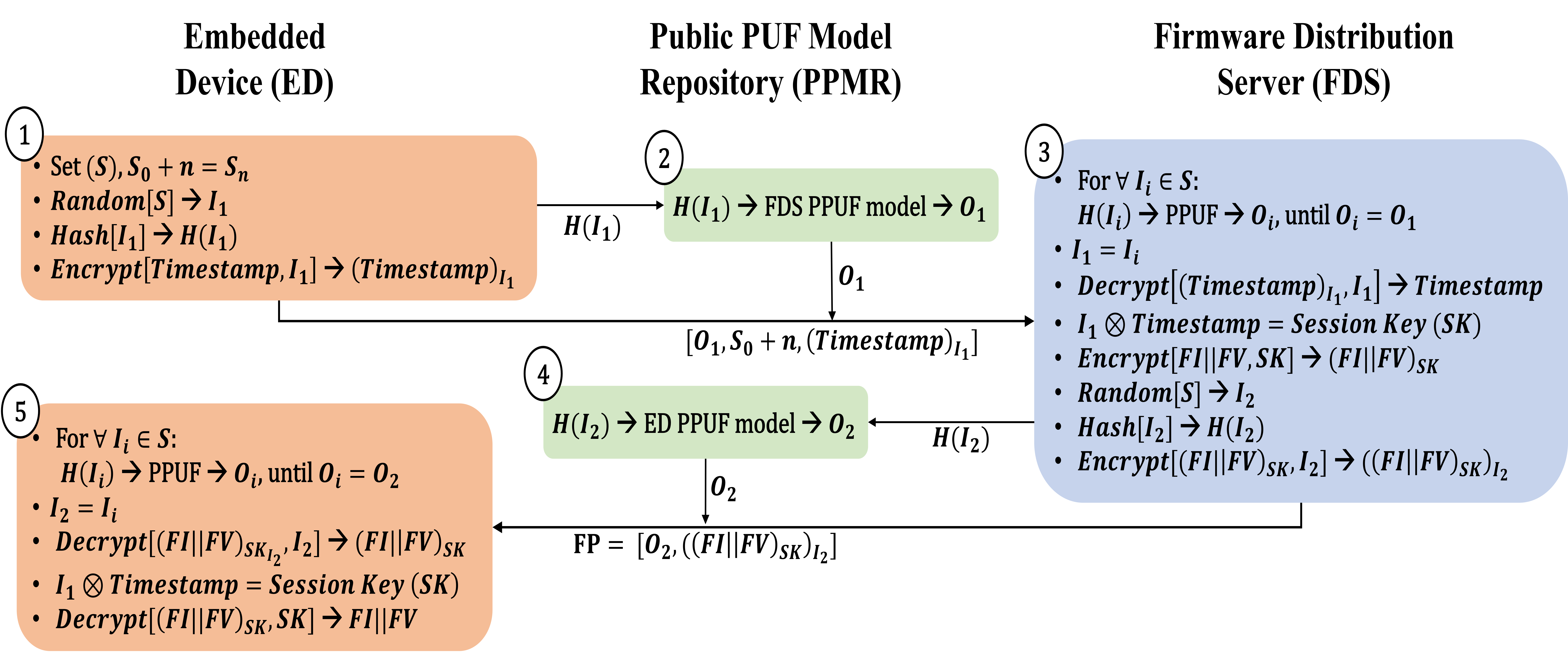}
    \caption{The proposed framework supports a communication protocol: an elaborate request-response handshake that provides security and modularity. The ED utilizes the software models stored on the Public PUF Model Repository (PPMR) to request a firmware update from the Firmware Distribution Server (FDS). In response, the FDS encrypts and forwards a firmware package to the device with the help of the PPMR.}
    \label{fig:protocol steps}
\end{figure*}

\subsection{Firmware Update Protocol Design} \label{ss:firmware update protocol design}

The steps of the proposed protocol can be seen in Fig. \ref{fig:protocol steps}. The firmware update protocol is divided into three main procedures: i) the ED requests a firmware update from the FDS with the help of the PPMR for creating the key (steps 1-2 in Fig. \ref{fig:protocol steps}), ii) the FDS prepares an encrypted package containing the firmware image and metadata that only the PPUF-enabled device can utilize (steps 3-4 in Fig. \ref{fig:protocol steps}), and iii) the device verifies the validity and originality of the firmware package's contents and then proceeds to decrypt and apply the new image (step 5 in Fig. \ref{fig:protocol steps}). The FDS is a server containing the firmware updates for the EDs while PPMR is a publicly accessible entity that holds the software model of every PPUF participating in the communication.

The client is responsible for initiating the firmware update process for several reasons: (1) this is common practice among Over-The-Air (OTA) update policies which are prevalent in the mobile computing and IoT areas  \cite{zigbeealliance2019,moran2019firmware}, (2) avoid service disruption at inconvenient times as well as evade  possible catastrophic results in industrial processes that might be  sensitive to slight process deviations, and (3) avoid having the server broadcasting firmware packages to the candidate devices which can overload and increase the bandwidth requirements of a network.

\subsubsection{Embedded Device Firmware Update Request}
In step (1), the ED requesting a firmware update will first pick a random element $(I_1)$ from a very large set of numbers $S$, using it as one of the encryption keys. Using $S_0+n$, we can compactly represent the set $S$, where $S_0$ is the initial number of the set and $n$ represents the offset.
This set of numbers can be as large as the system integrator's constraints in order to favor security requirements (by choosing a very large set) or faster firmware updates (smaller set). We use $S$ as an indicator to narrow down the search space for a valid challenge-response pair since the 
$2^{256}$ combinations of possible challenges would be extraordinarily slow.
Then, the element chosen $I_1$, is hashed ($H(I_1)$) and sent to the PPMR as a challenge to the FDS PPUF model. In step (2), the PPMR using the $H(I_1)$, challenges the FDS' PPUF model to create the response $O_1$.
In order to proceed to step (3), the response $O_1$ is sent to the FDS along with an encrypted $(Timestamp)_{I_1}$. The $Timestamp$ is used to satisfy \textit{SDR1}, and the compact representation of the set $S$, $S_0+n$.

The random element $I_1$ chosen to be the PPUF model challenge is hashed in order not to leak information to an attacker. Even if an attacker has access to the hashed challenge and the plaintext response of the PPMR, it is infeasible to get hold of $I_1$ due to the non-reversible nature of cryptographic hash functions. The only way to uncover the origins of $H(I_1)$ is to iterate throughout every element $I_i$ in set $S$, hash it, and send it to the correct PPMR model until $O_i = O_1$. This brute force approach demonstrates that $I_1$ can be used as a key to encrypt the timestamp since it is infeasible to complete this procedure during simulation.

\subsubsection{Encrypted Firmware Package Preparation}
In step (3), the FDS receives $[O_1, S_0+n, (Timestamp)_{I_1}]$ and uses its PPUF to determine which $I_1 \in S$, when hashed, produces $O_1$, i.e., it calculates $I_1$. Once this is completed, the FDS proceeds to decrypt the $(Timestamp)_{I_1}$ using $I_1$ and creates a temporary Session Key ($SK$), by using bitwise XOR between $I_1$ and the $Timestamp$. The FDS then uses the $SK$ to encrypt a concatenation of the firmware image $(FI)$ with the firmware version code $(FV)$, which includes best-before timestamps, vendor, and device-type identifiers, and software revision code. As per \textit{SDR6}, the $FI$ needs to be encrypted when transmitted. The encrypted $FI$ is accompanied by the aforementioned metadata in order to satisfy the need for protection against rollback attacks and avoid ED mismatches, as per \textit{SDR2} and \textit{SDR3}. Consequently, the FDS selects a random element $I_2$ from $S$ and sends its hash, $H(I_2)$, to the PPMR as a challenge to the ED PPUF model, which in turn will provide the appropriate response $O_2$ (step (4)). At the same time, the FDS uses $I_2$ to re-encrypt $(FI||FV)_{SK}$ to obtain $((FI||FV)_{SK})_{I_2}$. Finally, the server sends the complete firmware package to the ED, containing the firmware image as well as the necessary metadata alongside the response $O_2$ obtained from the PPMR.

The process to determine $I_i$ is performed completely on hardware. It is considerably faster than a simulating attacker due to the ESG. The firmware image is encrypted twice, a process often called \textit{signcryption}, in order to: (1) ensure that only the device requesting the firmware is able to decrypt it correctly and in a reasonable time by using its PPUF's model, and (2) for the ED to verify the firmware's origins. The ED can ensure that the firmware comes from the authenticated source because only the FDS can decrypt the firmware update request and recreate the $SK$ on its end during the set timeframe. This method satisfies both \textit{SDR4} and \textit{SDR5} since the firmware is digitally signed providing proof of its authenticity.
The process ensures that every firmware update package delivered to each device is encrypted in a way that only the indented device can decrypt it. By having chip-specific firmware update packages we gain significant security advantages. For example, an attacker might attempt to get access to a firmware package by creating clone devices in order to trick the FDS into creating a firmware package  \cite{karri2017physical}. However, due to the uniqueness of the PPUF any other device, even cloned ones, cannot decrypt the firmware package correctly and in a feasible manner. Also, this prevents an attacker from forwarding a firmware package to a different device. 
Chip-specific encryption provides to the integrator the ability not to rely on forgeable data such as serial numbers or MAC addresses for device identification.

\subsubsection{Firmware Update Verification and Unpacking}
The ED receives the encrypted firmware package $(FP)$ as part of step (5). First, the ED creates a new timestamp and compares it to the original timestamp created when the procedure was initialized. The new timestamp is used to create a timeframe in which a response will be accepted. This timing threshold has to be short enough so that only the FDS could reply in time. The two timestamps are compared, and if their difference is greater than the pre-set threshold, the $FP$ is considered potentially dangerous since it did not arrive in the expected timeframe. If the $FP$ is received in the specified timeframe, it then utilizes its PPUF to determine $I_2$ by iterating among all possible $H(I_2)$ within the set $S$ in order to find a matching response to $O_2$. $I_2$ is utilized as the key for removing the first layer of encryption. The second pass of decryption needs $SK$ as the decryption key. Since the ED already knows $I_1$ and the $Timestamp$, it can create the $SK$ on its end by using bitwise XOR between the two. The second decryption procedure reveals the firmware image, version, and other identifiers included in the package. These identifiers are utilized, as suggested by \textit{SDR2} and \textit{SDR3}, to confirm the firmware image's compatibility with the ED and check the accompanying timestamps.
If any of the above steps ends up in a signature or key mismatch, it is considered as a failed firmware update attempt, and the firmware package's data is discarded. If such events happen repeatedly in a short time, the device triggers a cooldown period that disables firmware updates. Until action is taken by the system administrator, the firmware updates stay disabled and the device continues its normal operation, which satisfies \textit{SDR7}.

\subsection{Framework Features and Characteristics}
\subsubsection{Modularity and Flexibility}
Cryptographic primitives can widely vary in terms of the provided security level and throughput. In our case, all of them are implemented on hardware, thus area and performance must be taken into consideration. The firmware update protocol in our framework is designed with modularity in mind; to allow the integrator to select the cryptographic primitives between different variants and types, i.e., encryption/decryption functions, hashing algorithms, and propagation delay-based PPUF. The framework allows for customization of the proposed solution by taking into consideration the system's constraints and requirements.
This is possible because the firmware update protocol does not rely on certain cryptographic implementations but rather on their functionality. As a result, hardware-implemented security primitives with the same functionality can be interchanged according to the implementation requirements of each deployment scenario. 

\subsubsection{Interoperability and Compatibility}
The proposed update procedure is agnostic to how firmware images are distributed. Firmware images can be delivered to EDs in a variety of ways over wired or wireless network protocols. The update mechanism can be adapted to operate in any specific case since it can work in conjunction with any data transmission mechanism or protocol. Adding to the compatibility of the approach is the fact that this mechanism can be used for broadcast deliveries of firmware, which is an important aspect of firmware update mechanisms for IoT \cite{moran2019firmware}. The same firmware package can be aimed at a multitude of devices while making sure that only compatible devices accept them since the package is chip-specific.
These characteristics make the proposed firmware update protocol appealing to cases that involve large-scale deployments of EDs.

\section{Security Analysis} \label{s:security Analysis}

To validate the SDRs set in Section \ref{ss:secure design requirements} and implemented into the proposed protocol of Section \ref{ss:firmware update protocol design}, first we describe the threat model to highlight the attacker's capabilities. Then, we analyze the proposed firmware update protocol for various threats to examine how these attacks are thwarted by design.

\subsection{Threat Model} \label{ss:Threat Model}
In this work, we consider the Dolev-Yao threat model where any communication channel between two parties is considered insecure \cite{dolevyao}. The Dolev-Yao adversary has complete control over the network, as an authenticated user, and is capable of injecting, eavesdropping, modifying, and blocking messages on the communication network \cite{do2019role}. Moreover, we assume that the malicious adversary can obtain any data stored in non-volatile memories of the devices such as hard disks and other long-term storage elements. Also, the adversary can access the PPMR by pinging the domain for information, e.g., send a challenge for a certain PPUF and get the corresponding simulated response. It must be noted that this process will always be significantly slower than performing this procedure directly on the hardware (due to the ESG). However, the adversary cannot mount implementation attacks\footnote{Implementation attacks include side-channel and fault analysis attacks, probing attacks, hardware reverse engineering and any combinations of them \cite{popp2009introduction}. They are preventable through hardware intrusion detection and fault detection mechanisms as well as anti-radiation coating.} against the devices, cannot reverse engineer the PUFs nor obtain intermediate variables stored in flip-flops or on-device volatile memory such as caches, RAM, and other temporary registers. Device-level countermeasures for these attacks are beyond the scope of this paper. The main goal of the attacker in our scenario is to get hold of the firmware binary plaintext. This allows the attacker to uncover functionality about the embedded device that may help in other types of attacks. It also allows creating ``trojan-ed'' firmware code by injecting malicious routines or backdoors in the binary and then tricking the device into accepting it as a legitimate one. In terms of computing capabilities, the PPMR and FDS are high-performance computing units capable of handling strenuous tasks. The attacker has access to similar hardware and comparable computational capabilities.

\subsection{Threat Modeling and Evaluation} \label{ss:security evaluation}
The firmware update process must be secured from rollback and impersonation attacks. Specifically, the ED should not be able to install obsolete firmware with known vulnerabilities. Impersonation attacks (masquerade, man-in-the-middle, spoofing) must also be addressed since the attacker can redirect traffic and inject malicious data into the insecure channel. The security objectives of the proposed protocol include: (1) ensuring the confidentiality of the firmware image, (2) making the ED able to authenticate its source, and (3) guaranteeing data integrity.

We perform threat modeling and evaluation using the STRIDE approach \cite{kohnfelder1999threats,khan2017stride}. STRIDE is an acronym for Spoofing, Tampering, Repudiation, Information Disclosure, Denial of Service, and Elevation of Privilege. Our choice is motivated by (1) STRIDE's systematic approach revealing threats in each system component, (2) taking into consideration all the required security properties such as authentication, authorization, confidentiality, integrity, and non-repudiation, and (3) the fact that STRIDE provides a clear view of the consequences of each component's vulnerabilities and how they affect the whole system. We evaluate the security of the proposed framework against the following threats by examining our SDRs and how they act as countermeasures.

\textbf{Rollback Firmware (Escalation of Privilege):} An attacker sends a firmware of a previous version. The firmware package, despite being valid, can re-introduce vulnerabilities fixed in newer firmware versions, aiming to take control over the device. In the proposed protocol, this is prevented by using timestamps (\textit{SDR1}); if the installed firmware's timestamp is newer than the firmware update presented to the device, the device will reject the update.

\textbf{Mismatched Firmware (Denial of Service):} An attacker sends an unmodified firmware image but for the wrong type and model of the device in order to ``brick" it. That firmware package is signed correctly by the manufacturer. In case the device mistakenly accepts the update, it could cause hardware malfunctions, expose security vulnerabilities, or even render the device inoperable. However, our protocol ensures that the firmware image is accompanied by several device-type identifiers and is encrypted in a chip-specific way (\textit{SDR2}). Thus, the ED will reject ineligible firmware packages.

\textbf{Obsolete Firmware Update (Spoofing):} This attack is specifically aimed at EDs that were either offline for some time before coming back online or EDs that were neglected and not updated for some time. An attacker targets a device like these and tries to update the device with an obsolete firmware image, newer than the one currently installed but not the latest released firmware version. If there is a known vulnerability in the provided firmware, it may allow an attacker to gain control over the device. Such attacks are prevented due to the \emph{best-before timestamps} in the firmware package's metadata (\textit{SDR3}). The device rejects firmware packages that have a best-before time smaller than the current Unix time.

\textbf{Redirection (Denial of Service/Spoofing):} An attacker redirects network traffic and aims to impersonate FDS. This may allow the attacker to update a device with a compromised version of the firmware image.
The firmware update procedure starts with an encrypted request sent from the ED to the FDS. To decrypt the request, gain access to the timestamp, and create the correct temporary session key in a reasonable time, the attacker has to have the FDS' actual PPUF due to the hardware uniqueness. Trying to break the protocol using brute force approaches means simulating all possible CRPs of the PPUF. This can take from days to years, depending on the integrator's design choices.
What is more, the device is configured to have a deadline for accepting updates after the update request has been issued. Therefore, even if the attacker manages to acquire the unique-per-session key, his/her response cannot be in time to be accepted as a valid firmware (\textit{SDR1} and \textit{SDR3}). This deadline-type of approach can protect the device against other more sophisticated attacks as well, as discussed in Section \ref{sss:MITM}. In case the adversary constructs a firmware package aiming to launch a guessing attack, \textit{SDR4} guarantees that every firmware package is signed with the temporary session key so that the device can verify its source.

\textbf{Unauthenticated Updates (All STRIDE threats):} An attacker tries to install a maliciously modified firmware on a device, through payload or metadata manipulation, to gain control of the device. Data manipulation of the firmware package during its transmission through the insecure channel alters the hash digest created by the recipient ED. The mismatching hash digests trigger a firmware update rejection due to corrupted data (\textit{SDR5} and \textit{SDR7}) and prevent unauthenticated updates.

\textbf{Reverse Engineering of Firmware Image (All STRIDE threats):} An attacker intercepts the firmware package during its transmission over the insecure channel. To prepare an attack, the firmware package is decomposed, decrypted, and analyzed, allowing the attacker to perform reverse engineering of the firmware image in order to exploit possible vulnerabilities or introduce new modified subroutines. The firmware package includes an encrypted version of the image, with the key being created by the hardware of the recipient ED, thus preventing reverse engineering (\textit{SDR6}).

\subsection{Attack Vectors and Security Discussion} \label{ss:Attack Vectors}
As discussed in Section \ref{ss:security evaluation}, our proposed methodology is inherently and by-design resilient to a variety of attack vectors due to adhering to the SDRs set in Section \ref{ss:secure design requirements}. In this part, we further elaborate on the security of the proposed methodology extensively. Specifically, we focus on attack scenarios in which the secure-by-design characteristic of our framework might not be adequate and the attack may not be thwarted in a predetermined fashion such as in Section \ref{ss:security evaluation}.

\subsubsection{Attack Scenario 1:} \label{sss:Dictionary}
In the scenario in which a malicious adversary aims to accumulate previously used keys (e.g., CRPs) in a dictionary-type of attack, he/she will gain an advantage over the legitimate participating parties.  The attacker can amass information about the used CRPs  using various methods. For example, by eavesdropping on the traffic of the network, or by continuously pinging the PPMR with consecutive challenge strings to make a complete dictionary of CRPs. This kind of knowledge can be leveraged by an attacker to break the protocol by searching in the pre-computed keys when encountering a known CRP during a firmware update. However, we argue that our methodology is inherently resistant to this kind of attack. 

In our threat model, we consider an attacker that has the same computational capabilities as the FDS. However, let us assume an attacker with infinite processing power who can instantly compute every CRP combination required to have a full dictionary of every possible CRP. Considering that our dPPUF utilizes $256$-bit strings as Challenges/Responses and the fact that a CRP needs to be stored in pairs, a fully pre-computed dictionary would require $2*2^{256}$ bits of storage. Therefore, an attacker with finite storage capabilities but infinite processing power cannot store this amount of information, which sums up to $\approx 2.9 * 10^{61}$ petabytes.

The attacker could try to create a dictionary with a more manageable size, anticipating that one of the pre-computed CRPs will be the same as the one utilized in a firmware update that is currently under malicious monitoring. Taking into consideration the random challenge generation procedure, the chance that the attacker will have pre-computed a certain CRP, which happened to be the random challenge chosen for the firmware update, is $1/{2^{256}}$ (which is $\approx 2.26^{-72}$). To increase the chance of having the random challenge pre-computed, the attacker has to pre-compute large amounts of CRPs. By using Eq. \ref{eq:chanceToFindKey}, we can calculate the number of CRPs that have to be pre-computed for the attacker’s dictionary. The more dictionary entries, the higher the probability that one of them will contain the CRP currently in use.
\begin{equation}
    \label{eq:chanceToFindKey}
    \frac{X}{2^{256}} = C
\end{equation}
\noindent where $X$ is the amount of CRPs pre-computed and $C$ is the probability that one of them is the correct CRP. For example, for the attacker to have $C=1\%$ probability to have the needed CRP, a dictionary with $X=C*2^{256}$ is required. This amounts to a pre-computed dictionary $X\approx 1.16*10^{75}$ entries, that is, $\approx 10^{75}*256*2$ bits or $6.4 * 10^{61}$ petabytes. Therefore, the size of the pre-computed dictionary necessary to allow the attacker to have a strong probability of having the needed CRP is clearly prohibitive.

\subsubsection{Attack Scenario 2:}\label{sss:MITM}
The attacker can intervene with a complex procedure in a man-in-the-middle (MITM) type of attack and create an infected firmware package to pass as the  legitimate firmware update. To achieve this, the attacker will have to outrun the server. Specifically, the attacker will have to access messages between the ED and FDS in order to intercept the firmware package. The attacker will decompose the $FP$ and then reconstruct it using an infected firmware image. This procedure has to be done in a specific amount of time since the embedded device is not accepting firmware packages after a set amount of time (see Section \ref{ss:security evaluation} -- Redirection).

In order to effectively compare the workload of the FDS and the attacker, we disregard any network delays and assume that they will be the same on both sides. Also, we assume that the required timing for decryption and encryption is  the same for both parties when encrypting/decrypting the same amount of data. Finally, even though the server has dedicated hardware accelerators for encryption and hashing, we  assume that the attacker has the same computational capabilities as the FDS. Therefore, procedures such as hashing and encryption/decryption will take the same amount of computational effort on both sides.

Now let us examine the steps in which the attacker might take to perform a MITM attack. First, the attacker will have to intercept the first message from the ED to the FDS in order to acquire the subset $S_0+n$. Utilizing this information, the attacker can then create a dictionary of size $n$ by hashing all possible combinations in the subset $(n* t_{hash})$, and check which one matches $H(I_1)$. Assuming the attacker can utilize a hash table indexing structure, each search query will take $t_{hash}$ time. After finding $I_1$, the attacker will have to decrypt the $Timestamp$ $(t_{dec_1})$ and then let the ED request reach the FDS. After the FDS sends $H(I_2)$ to the PPMR, the attacker can search his database again for $I_2$ $(t_{hash})$, and let the FDS send out the $FP$. The attacker, now having access to $I_1$, $I_2$, $Timestamp$ and in extension $SK$, can try to decrypt the $FP$ using $I_2$ $(t_{dec_2})$ and then $SK$ $(t_{dec_2})$. To load the malicious $FI$ in a way that will fool the ED into accepting it, the attacker has to re-encrypt $FI||FV$ twice using $SK$ $(t_{enc_2}=t_{dec_2})$ and $I_2$ $(t_{enc_2}=t_{dec_2})$, respectively. The execution time of this procedure is summarized in Eq.  \ref{eq:AttackerSide} taking into account all appropriate simplifications.
\begin{equation}
    \label{eq:AttackerSide}
    \textit{Attacker Execution Time} = (n+2)*t_{hash} + t_{dec_1} + 4*t_{dec_2}
\end{equation}

The server has to perform the corresponding associated actions (step 3 of Figure \ref{fig:protocol steps}) to prepare the $FP$ for the ED. First, the FDS will have to utilize its dPPUF to loop between $S_0+n$ in order to find $I_1$ $(n*(t_{hash}+dPPUF_{gen}))$. After finding $I1$, the FDS has to decrypt the timestamp $(t_{dec_1})$ and create the $SK$ which is considered negligible since it's a simple bitwise-XOR operation. The server then encrypts the $FI||FV$ using $SK$  $(t_{enc_2})$ and hashes $I_2$ $(t_{hash})$ to be sent to the PPMR. While waiting for the PPMR's response, the server re-encrypts $(FI||FV)_{SK}$ with $I_2$ $(t_{enc_2})$. The result, combined with $O_2$, completes the $FP$ and is transferred to the ED. If we apply the appropriate simplifications and disregard the term $(n*dPPUF_{gen})$ due to being a very fast procedure in the order of picoseconds  \cite{potkonjak2011differential}, we can express the server's execution time using Eq.  \ref{eq:ServerSide}.
\begin{equation}
    \label{eq:ServerSide}
    \textit{Server Execution Time} = (n+1)*t_{hash} + t_{dec_1} + 2*t_{dec_2}
\end{equation}

In order for the firmware package delivery deadline set by the ED to be effective, the \textit{Server Execution Time} has to be significantly smaller than the \textit{Attacker Execution Time}. We can effectively compare Eq. \ref{eq:AttackerSide} and Eq. \ref{eq:ServerSide} by eliminating the common terms found in each formula and compare the server's advantage to the attacker.
\begin{equation}
    \label{eq:InequalityAdv}
    t_{hash} + 2*t_{dec_2} > t_{dec_2}
\end{equation}
As seen in Eq. \ref{eq:InequalityAdv}, even with all the relaxations in our assumptions, the attacker will always have to do at least more than twice the work required by the FDS in the same time frame. Therefore, the ED's deadline can be configured accordingly to be long enough to allow the FDS enough time to deliver the $FP$, but short enough to block any attacker's attempt to create and forward a forged $FP$.

\section{Experimental Setup and Results}\label{s:experiments}

\begin{figure}[t]
    \centering
    \includegraphics[scale=0.1]{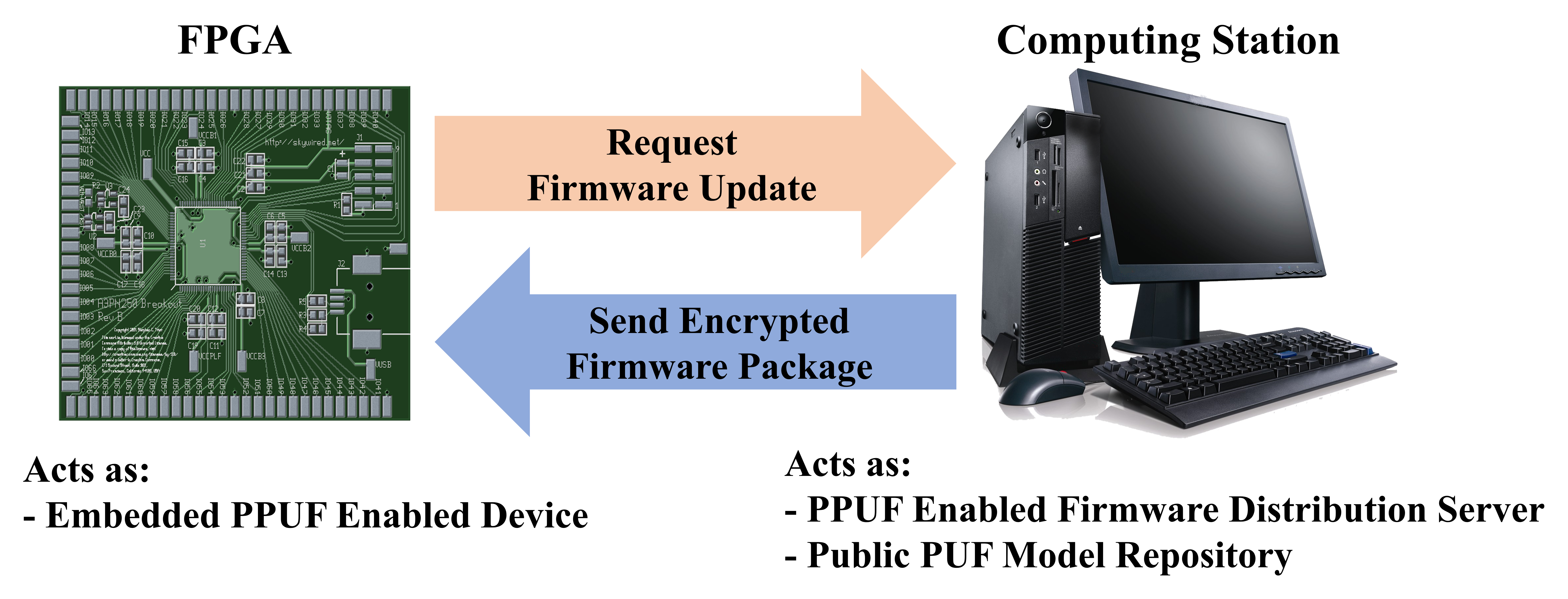}
    \caption{Experimental/evaluation setup. The hardware-implemented security primitives are developed on a FPGA in order to emulate a PPUF-enabled device. The servers and firmware packaging procedure are carried out on a computer which is connected to the FPGA through a serial cable.}
    \label{fig:exp setup}
\end{figure}

The experimental setup used to validate and evaluate the effectiveness of our proposed methodology is shown in Fig. \ref{fig:exp setup}. A Xilinx Kintex7 FPGA emulates the PPUF-enabled ED while a connected computer acts as both the PPMR and FDS.
The computer is a $64$-bit machine with $3.2GHz$ Intel Core $i5-4460$ quad-core processor, with $8GB$ RAM, running on Windows 10, and connects to the FPGA through a dedicated serial port. The FPGA initiates a firmware update request, while the computer responds from the PPMR, packages the firmware image, and sends it to the FPGA through the serial port. The FPGA, then, decrypts and validates the firmware. During this procedure, the FPGA logs and reports the elapsed time back to the computer.

We use Python's library \textit{pycryptodome} to implement the software for encryption and hashing on the computer's side. The simulation model of the dPPUF, present in the PPMR, is developed in \textit{C++} for demonstration purposes. The serial cable, connecting the computer and the FPGA, uses a $115200$ baud rate, $8$ data bits, no parity, $1$ stop bit, and no flow control. The firmware update request contains $20$ bits for the compact representation of the set $S$, $128$ bits for the encrypted timestamp, and $256$ bits for the response $O_1$. The additional metadata accompanying the encrypted firmware image, the computer sends to the FPGA, amounts to $256$ and $10$ bits for the response $O_2$ and firmware version code $FV$, respectively. Both firmware update request and $(FI||FV)$ are accompanied by a $256$-bit or $512$-bit hash digest, depending on the hash functions used (SHA-256 or SHA-3), utilized as a checksum value.

In order to emulate manufacturing variation on the dPPUF's design, we randomize each gate's switching delays on the FPGA (shown in Fig. \ref{fig:dppuf}). The design consists of multiple alternating layers of boosters (2-input XOR gates) and repressers (NAND-based circuit), using the default configuration suggested in  \cite{potkonjak2011differential}. 
However, we increase the width of the dPPUF from 64 bits to 256 to enhance the simulation complexity, hence providing better security guarantees.
To validate the entropy of the design, we conduct test runs of the CRP functionality of the dPPUF using $10k$ input vectors. Using the Strict Avalanche Criterion (SAC), i.e., the probability that an output bit will switch for inputs of hamming distances 1,
we test the randomness of the output and its ability to be correlated to the input. The results show that the design is adequately random, as suggested in  \cite{potkonjak2011differential}, showing an average switching probability of 0.3425. The set $S$, which the protocol requires as a seed of random keys, is set to a range of $10^6$.

\begin{table}[t]
\centering
\caption{Hardware resources required by components implemented on the FPGA}
\label{tab:component resources}
\begin{tabular}{||c||c|c||}
\hline \hline
\textbf{Components} & \textbf{LUT (\#)} & \textbf{FF (\#)} \\ \hline \hline
\textbf{dPPUF} & 766 & 275 \\ \hline
\textbf{SIMON} & 275 & 201 \\ \hline
\textbf{SHA-256} & 1330 & 753 \\ \hline
\textbf{Twofish} & 1272 & 690 \\ \hline
\textbf{AES-GCM} & 2671 & 1568 \\ \hline
\textbf{SHA-3 (Keccak)} & 2605 & 2244 \\
\hline \hline
\end{tabular}
\end{table}

\begin{table}[t]
\centering
\caption{Embedded Device Configurations and Total Resource Requirements}
\label{tab:setups comps}
\begin{tabular}{||c||c|c|c|c|c||}
\hline \hline
\multirow{2}{*}{\textbf{Configurations}} & \multirow{2}{*}{\textbf{PPUF}} & \multirow{2}{*}{\textbf{Encryption}} & \multirow{2}{*}{\textbf{Hashing}} & \multicolumn{2}{c||}{\textbf{Total Resources}} \\ \cline{5-6}
 & & & & \textbf{LUT (\#)} & \textbf{FF (\#)} \\
\hline \hline
\textbf{Lightweight} & \multirow{3}{*}{dPPUF} & SIMON & \multirow{2}{*}{SHA-256} & 2366 & 1210\\ \cline{1-1} \cline{3-3} \cline{5-6}
\textbf{Midweight} &  & Twofish & & 3313 & 1699 \\ \cline{1-1} \cline{3-4} \cline{5-6}
\textbf{Heavyweight} &  & AES-GCM & SHA-3 & 5894 & 4068\\ 
\hline \hline
\end{tabular} 
\end{table}

The FPGA emulates the dPPUF-enabled device in order to better represent a real scenario. To achieve this, both dPPUF and the corresponding security primitives are implemented on hardware using VHDL. The FPGA clocks at $100MHz$ across all components to represent a resource-constrained device found in IoT environments. Using this configuration, we can accurately monitor the overhead and performance of the unpacking process in order to demonstrate the viability of our proposed methodology. The hardware-implemented cryptographic components used are presented in Table \ref{tab:component resources} alongside their requirements in hardware resources of the FPGA. We utilize a $256$-bit dPPUF, two different cryptographic hash functions, and three different encryption/decryption modules. Specifically, we choose SHA-256 and SHA-3 ($512$-bit version) as the hash functions to be tested. For encryption/decryption, SIMON ($64$-bit block size and $128$-bit key), Twofish ($128$-bit), and AES-GCM ($128$-bit) are chosen. This selection of primitives provides a comprehensive view of how state-of-the-art cryptographic functions of different security capabilities, performance, and area requirements can be integrated into the proposed framework.

To present the flexibility and modularity of our design, we utilize three different use cases representing different EDs. The cases' names reflect their relative hardware resource utilization, hence the names \textit{Lightweight}, \textit{Midweight}, and \textit{Heavyweight}. These scenarios showcase three dPPUF-enabled devices, unique in constraints and requirements in terms of security guarantees and resource utilization. The range from \textit{Lightweight} to \textit{Heavyweight} indicates the trade-off between resources and performance plus security level. The \textit{Lightweight} case is the least expensive in hardware resources and also lowest in performance and security. On the contrary, the \textit{Heavyweight} case is the fastest and most secure configuration with the largest hardware overhead.
These configurations are created with existing IoT devices in mind. For example, the \textit{Lightweight} configuration can be used in ultra-low-power devices such as battery-operated smart sensors. The \textit{Midweight} configuration is suitable for more capable IoT devices such as microprocessor-enabled smart appliances (e.g., smart WiFi light bulbs), while the \textit{Heavyweight} configuration is designed for more heavy-duty devices such as PLCs.
Table \ref{tab:setups comps} shows the hardware utilization by Xilinx Vivado Design Suite 2019.2 after the default optimizations occur (hence the disparity between Table I and Table II). The plug-and-play functionality of the framework enables the easy evaluation of different combinations. The presented components and modules are indicative and do not restrict the device integrator's selection if the components are of similar functionality.

\begin{table}[t]
\centering
\caption{Time to complete the update procedure in seconds}
\label{tab:fw sizes}
\begin{tabular}{||c||c|c|c||}
\hline \hline
\textbf{\begin{tabular}[c]{@{}c@{}}Configuration/\\ Firmware Image\end{tabular}} & \textbf{\begin{tabular}[c]{@{}c@{}}Sercos III\\ $233 kB$\end{tabular}} & \textbf{\begin{tabular}[c]{@{}c@{}}Zelio Logic\\ $323 kB$\end{tabular}} & \textbf{\begin{tabular}[c]{@{}c@{}}Modicon\\ $1183 kB$\end{tabular}} \\ \hline \hline
\textbf{Lightweight} & 0.3407 & 0.4722 & 1.7296 \\ \hline
\textbf{Midweight} & 0.3356 & 0.4653 & 1.7041 \\ \hline
\textbf{Heavyweight} & 0.3338 & 0.4627 & 1.6946\\ \hline \hline
\end{tabular} 
\end{table}

In addition to the different component configurations, we utilize three commercial firmware files (acquired from the vendors' websites) for testing our framework. We selected images for constrained embedded systems deployed in environments such as ICS and IIoT: these include a Sercos III field bus interface module, a Zelio Logic SR2/SR3 smart relay, and a Modicon M258 logic controller. We match these firmware images with all the hardware configurations to evaluate the performance of the framework and examine the impact of hardware selection on the overall performance of the device during the firmware update procedure. Table \ref{tab:fw sizes} shows the results of these experiments.
The updating procedure, depending on the firmware size, can take between $0.3$ to $1.72$ seconds. The performance of the proposed approach provides significant improvements over traditional updating procedures that, in some cases, can take hours to complete \cite{houstonchr}. Observing these configurations and their respective times to complete the updating procedure, we can conclude that performance is relative to resource usage, e.g., the \textit{Heavyweight} configuration outperforms the less resource-heavy configurations. Also, the most resource-demanding cryptomodules (AES-GCM, SHA-3) provide the best security guarantees among the implemented encryption and hash functions.
\section{Related work and comparison with proposed approach}\label{s:related work}

IoT manufacturers often provide the firmware images required for device operation online via their website. It has been shown that web crawlers can search and automatically acquire images of industrial equipment available in the market \cite{costin2014large}. Firmware images can also be acquired through physical access, by forcing memory dumps of onboard memory modules \cite{firmtaxon}. Access to firmware files and code can reveal device functionality. Firmware modification can have severe repercussions on the system operation, thus firmware is typically an alluring target for malicious actors. The severity of firmware attacks, specifically for the ICS and IIoT domain, has been shown in literature for devices such as programmable logic controllers and protection relays \cite{basnight2013firmware, konstantinou2015impact}. Attacks have also been targeting other generic EDs such as printers, cameras, and network switches \cite{costin2014large, cui2013firmware}. The need for firmware security measures is now more urgent than ever.

To the best of our knowledge, this is one of the first efforts to leverage PPUF-based end-to-end encryption to implement \emph{secure} authentication protocols for firmware updates. Our PPUF-based framework preserves firmware integrity and provides security from various threats (as discussed in Section \ref{ss:security evaluation}) not considered by existing techniques.
Our flexible framework demonstrates the ability to be reconfigured in terms of hardware components while also giving the ability to EDs to be deployed in-field without any secure enrollment phase or hardcoded secrets within the ED's non-volatile memory.
Recent work has shown that PPUFs alongside other security primitives implemented on hardware  could be utilized for secure firmware delivery to in-field devices \cite{falas2019hardware}. However, the approach is not end-to-end secure as it lacks the communication handshake required for two-way authentication in a truly public-key -based fashion. Therefore, the server is blind to who the firmware package recipient. Also, the updating procedure must be initiated from the firmware vendor's side which is not considered good practice in OTA updates \cite{zigbeealliance2019,moran2019firmware}. The work in \cite{falas2019hardwarespringer} improves the firmware updating procedure proposed in  \cite{falas2019hardware} by introducing digital signatures, which allow the device to better authenticate the manufacturer. However, it still suffers from the aforementioned drawbacks.
Similar approaches, utilizing different PUFs and other cryptographic primitives implemented in hardware, are only authentication-oriented protocols, such as  \cite{aysu2015end, che2017privacy}. We present a comparison between the aforementioned methodologies with our approach in Table \ref{tab:prev work}. To perform a meaningful comparison with respect to performance, we isolate the authentication part of our framework.
We assume that the authentication procedure takes place from step (1) to step (5), excluding the firmware image's encryption and decryption procedures. Observe that the overhead hardware of our \textit{Lightweight} and \textit{Heavyweight} configuration reported in Table \ref{tab:prev work} includes the LUT and FF for encryption/decryption. Our framework's flexibility to consider different configurations allows for comparison with both of the aforementioned related works.

\begin{table}[t]
\centering
\caption{Comparison with related authentication protocols}
\label{tab:prev work}
\begin{tabular}{||c||c|c||c||}
\hline
\hline
\multirow{2}{*}{\centering{\bf Method}} & \multicolumn{2}{c||}{\bf Area Overhead} & \multirow{2}{*}{\centering{\bf Performance}} \\
 \cline{2-3}
  & \textbf{LUT (\#)}    &  \textbf{FF (\#)} & \\ \hline \hline
{\cite{aysu2015end}}    & 3543      & 1275 &     0.061 sec \\ \hline
\textbf{Lightweight} & \textbf{2366} & \textbf{1210} & \textbf{0.064} sec\\
\hline \hline
{\cite{falas2019hardwarespringer}}    & 3183      & 3981 &     0.072 sec \\ \hline
\textbf{Midweight} & \textbf{3313} & \textbf{1699} & \textbf{0.029} sec\\
\hline \hline
{\cite{che2017privacy}} & 6038      & 1724   &   1.25 sec \\ \hline
\textbf{Heavyweight} & \textbf{5894} & \textbf{4068} & \textbf{0.015 sec}\\
\hline \hline
\end{tabular}
\vspace{-0.1cm}
\end{table}

In \cite{aysu2015end}, the authors utilize a combination of components similar to our \textit{Lightweight} setup, including the SIMON encryption core. However, their methodology does not take data integrity into account, therefore it does not utilize any hash functions. As shown in Table \ref{tab:prev work}, our framework performs as well as the one proposed by the authors. However, our implementation requires much fewer resources on the FPGA, even though the PUF used by \cite{aysu2015end} is an SRAM PUF, a much less resource-heavy component than the dPPUF.
The authors in \cite{falas2019hardwarespringer}, utilize a similar firmware update procedure using AES-GCM, SHA-256, an RSA core, and a dPPUF. While \cite{falas2019hardwarespringer} addresses data integrity issues of previous works in the area (e.g., \cite{aysu2015end},  \cite{che2017privacy}), it still lacks an end-to-end secure framework concept and demonstrates several shortcomings as explained in the previous paragraph.
In \cite{che2017privacy}, the authors demonstrate an authentication protocol that provides confidentiality and mutual authentication, without addressing data integrity issues, by utilizing a hardware-embedded path delay (HELP) PUF. The PUF derives randomness from path delay variance within a hardware implementation of AES. The work shows comparable resource usage to our \textit{Heavyweight} configuration by requiring almost the same amount of LUTs but significantly fewer FFs. Our implementation, however, performs better than the mechanism proposed by \cite{che2017privacy} by requiring orders of magnitude less time to complete the authentication procedure.

Other related, authentication-oriented works, such as \cite{haji2019public,wallrabenstein2016practical, braeken2018puf,chatterjee2018building,puf2019iomt}, cannot be compared directly to our work due to the lack of adequate quantitative data. In addition, such approaches have fundamental differences with our proposed concept and exhibit several drawbacks when compared to our methodology. They require an initial setup step in a secure environment. This initial setup often assumes a trusted third-party server that coordinates the handshakes and distributes keys over a secure channel before the ED's deployment. Our framework alleviates the need for an enrollment phase (on top of PUF hardware characterization) and the necessity for a secure channel. While the argument that the PPMR is a third-party server is valid, we do not rely on its trustworthiness nor we assume that its contents are secret. The PPMR's contents are publicly available.
\section{Conclusions and Future Work}\label{s:conclusions}

    In this paper, we develop a flexible and secure firmware update framework that is suitable for the diverse device-space of an IoT system. This framework's main contributions are adhoc device deployment without pre-installed security keys, fast firmware updates to remote devices,  chip-specific firmware package encryption,  ensure confidentiality and authenticity of the firmware binary through the use of PPUFs and other hardware-implemented cryptographic primitives, and an end-to-end approach that beyond authentication takes data integrity into account. Our experimental setup demonstrates the flexibility and security of our approach by selecting alternative underlying security primitives and testing the communication protocol against several attack vectors. A proof-off-concept implementation with commercial-of-the-shelf embedded devices' firmware images verifies the practicality of the approach.

   As future research, the proposed methodology will be examined under large-scale embedded device deployment scenarios that employ peer-to-peer network connectivity. For example, a swarm of IoT devices could utilize this protocol to perform peer-to-peer firmware delivery and authentication. Another direction, of our future work, is applying the proposed framework in specific application domains such as autonomous vehicles, where the protocol would be used to update the vehicles through a vehicle-to-infrastructure communication channel.

\section*{Acknowledgment}\label{s:acknowledgement}
This work has been supported by the EU Horizon 2020 research and innovation programme under grant agreement No 739551 (KIOS CoE) and from the Government of the Republic of Cyprus through the Directorate General for European Programmes, Coordination and Development. Partial support was also provided by the Woodrow W. Everett, Jr. SCEEE Development Fund in cooperation with the Southeastern Association of Electrical Engineering Department Heads.

\bibliographystyle{ACM-Reference-Format}
\bibliography{sources}


\begin{thebibliography}{35}


\ifx \showCODEN    \undefined \def \showCODEN     #1{\unskip}     \fi
\ifx \showDOI      \undefined \def \showDOI       #1{#1}\fi
\ifx \showISBNx    \undefined \def \showISBNx     #1{\unskip}     \fi
\ifx \showISBNxiii \undefined \def \showISBNxiii  #1{\unskip}     \fi
\ifx \showISSN     \undefined \def \showISSN      #1{\unskip}     \fi
\ifx \showLCCN     \undefined \def \showLCCN      #1{\unskip}     \fi
\ifx \shownote     \undefined \def \shownote      #1{#1}          \fi
\ifx \showarticletitle \undefined \def \showarticletitle #1{#1}   \fi
\ifx \showURL      \undefined \def \showURL       {\relax}        \fi
\providecommand\bibfield[2]{#2}
\providecommand\bibinfo[2]{#2}
\providecommand\natexlab[1]{#1}
\providecommand\showeprint[2][]{arXiv:#2}

\bibitem[\protect\citeauthoryear{Akhundov, Sluis, Hamdioui, and
  Taouil}{Akhundov et~al\mbox{.}}{2019}]%
        {haji2019public}
\bibfield{author}{\bibinfo{person}{Haji Akhundov}, \bibinfo{person}{Erik
  Sluis}, \bibinfo{person}{Said Hamdioui}, {and} \bibinfo{person}{M. Taouil}.}
  \bibinfo{year}{2019}\natexlab{}.
\newblock \showarticletitle{Public-Key Based Authentication Architecture for
  IoT Devices Using PUF}. In \bibinfo{booktitle}{\emph{CSEIT}}.
  \bibinfo{pages}{353--371}.
\newblock


\bibitem[\protect\citeauthoryear{Alliance}{Alliance}{2019}]%
        {zigbeealliance2019}
\bibfield{author}{\bibinfo{person}{Zigbee Alliance}.}
  \bibinfo{year}{2019}\natexlab{}.
\newblock \bibinfo{title}{Zigbee Cluster Library}.
\newblock
\newblock
\urldef\tempurl%
\url{https://zigbeealliance.org/developer_resources/zigbee-cluster-library/}
\showURL{%
\tempurl}


\bibitem[\protect\citeauthoryear{Aysu, Gulcan, Moriyama, Schaumont, and
  Yung}{Aysu et~al\mbox{.}}{2015}]%
        {aysu2015end}
\bibfield{author}{\bibinfo{person}{Aydin Aysu}, \bibinfo{person}{Ege Gulcan},
  \bibinfo{person}{Daisuke Moriyama}, \bibinfo{person}{Patrick Schaumont},
  {and} \bibinfo{person}{Moti Yung}.} \bibinfo{year}{2015}\natexlab{}.
\newblock \showarticletitle{End-to-end design of a PUF-based privacy preserving
  authentication protocol}. In \bibinfo{booktitle}{\emph{International Workshop
  on Cryptographic Hardware and Embedded Systems}}. Springer,
  \bibinfo{pages}{556--576}.
\newblock


\bibitem[\protect\citeauthoryear{Basnight, Butts, Lopez~Jr, and Dube}{Basnight
  et~al\mbox{.}}{2013}]%
        {basnight2013firmware}
\bibfield{author}{\bibinfo{person}{Zachry Basnight}, \bibinfo{person}{Jonathan
  Butts}, \bibinfo{person}{Juan Lopez~Jr}, {and} \bibinfo{person}{Thomas
  Dube}.} \bibinfo{year}{2013}\natexlab{}.
\newblock \showarticletitle{Firmware modification attacks on programmable logic
  controllers}.
\newblock \bibinfo{journal}{\emph{International Journal of Critical
  Infrastructure Protection}} \bibinfo{volume}{6}, \bibinfo{number}{2}
  (\bibinfo{year}{2013}), \bibinfo{pages}{76--84}.
\newblock


\bibitem[\protect\citeauthoryear{Braeken}{Braeken}{2018}]%
        {braeken2018puf}
\bibfield{author}{\bibinfo{person}{An Braeken}.}
  \bibinfo{year}{2018}\natexlab{}.
\newblock \showarticletitle{PUF based authentication protocol for IoT}.
\newblock \bibinfo{journal}{\emph{Symmetry}} \bibinfo{volume}{10},
  \bibinfo{number}{8} (\bibinfo{year}{2018}), \bibinfo{pages}{352}.
\newblock


\bibitem[\protect\citeauthoryear{Chatterjee, Govindan, Sadhukhan, Mukhopadhyay,
  Chakraborty, Mahata, and Prabhu}{Chatterjee et~al\mbox{.}}{2018}]%
        {chatterjee2018building}
\bibfield{author}{\bibinfo{person}{Urbi Chatterjee}, \bibinfo{person}{Vidya
  Govindan}, \bibinfo{person}{Rajat Sadhukhan}, \bibinfo{person}{Debdeep
  Mukhopadhyay}, \bibinfo{person}{Rajat~Subhra Chakraborty},
  \bibinfo{person}{Debashis Mahata}, {and} \bibinfo{person}{Mukesh~M Prabhu}.}
  \bibinfo{year}{2018}\natexlab{}.
\newblock \showarticletitle{Building PUF based authentication and key exchange
  protocol for IoT without explicit CRPs in verifier database}.
\newblock \bibinfo{journal}{\emph{IEEE Transactions on Dependable and Secure
  Computing}} \bibinfo{volume}{16}, \bibinfo{number}{3} (\bibinfo{year}{2018}),
  \bibinfo{pages}{424--437}.
\newblock


\bibitem[\protect\citeauthoryear{Che, Martin, Pocklassery, Kajuluri, Saqib, and
  Plusquellic}{Che et~al\mbox{.}}{2017}]%
        {che2017privacy}
\bibfield{author}{\bibinfo{person}{Wenjie Che}, \bibinfo{person}{Mitchell
  Martin}, \bibinfo{person}{Goutham Pocklassery}, \bibinfo{person}{Venkata~K
  Kajuluri}, \bibinfo{person}{Fareena Saqib}, {and} \bibinfo{person}{Jim
  Plusquellic}.} \bibinfo{year}{2017}\natexlab{}.
\newblock \showarticletitle{A privacy-preserving, mutual PUF-based
  authentication protocol}.
\newblock \bibinfo{journal}{\emph{Cryptography}} \bibinfo{volume}{1},
  \bibinfo{number}{1} (\bibinfo{year}{2017}), \bibinfo{pages}{3}.
\newblock


\bibitem[\protect\citeauthoryear{Colombier, Bossuet, Fischer, and
  H{\'e}ly}{Colombier et~al\mbox{.}}{2017}]%
        {colombier2017key}
\bibfield{author}{\bibinfo{person}{Brice Colombier}, \bibinfo{person}{Lilian
  Bossuet}, \bibinfo{person}{Viktor Fischer}, {and} \bibinfo{person}{David
  H{\'e}ly}.} \bibinfo{year}{2017}\natexlab{}.
\newblock \showarticletitle{Key reconciliation protocols for error correction
  of silicon PUF responses}.
\newblock \bibinfo{journal}{\emph{IEEE Transactions on Information Forensics
  and Security}} \bibinfo{volume}{12}, \bibinfo{number}{8}
  (\bibinfo{year}{2017}), \bibinfo{pages}{1988--2002}.
\newblock


\bibitem[\protect\citeauthoryear{{Common Vulnerabilities and Exposures
  (CVE\textregistered) List, The MITRE Corporation}}{{Common Vulnerabilities
  and Exposures (CVE\textregistered) List, The MITRE Corporation}}{2017}]%
        {CVE-2017-5698}
\bibfield{author}{\bibinfo{person}{{Common Vulnerabilities and Exposures
  (CVE\textregistered) List, The MITRE Corporation}}.}
  \bibinfo{year}{2017}\natexlab{}.
\newblock \bibinfo{title}{{CVE-2017-5698}}.
\newblock
\newblock
\urldef\tempurl%
\url{https://cve.mitre.org/cgi-bin/cvename.cgi?name=CVE-2017-5698}
\showURL{%
\tempurl}


\bibitem[\protect\citeauthoryear{Costin, Zaddach, Francillon, and
  Balzarotti}{Costin et~al\mbox{.}}{2014}]%
        {costin2014large}
\bibfield{author}{\bibinfo{person}{Andrei Costin}, \bibinfo{person}{Jonas
  Zaddach}, \bibinfo{person}{Aur{\'e}lien Francillon}, {and}
  \bibinfo{person}{Davide Balzarotti}.} \bibinfo{year}{2014}\natexlab{}.
\newblock \showarticletitle{A large-scale analysis of the security of embedded
  firmwares}. In \bibinfo{booktitle}{\emph{23rd $\{$USENIX$\}$ Security
  Symposium ($\{$USENIX$\}$ Security 14)}}. \bibinfo{pages}{95--110}.
\newblock


\bibitem[\protect\citeauthoryear{Cui, Costello, and Stolfo}{Cui
  et~al\mbox{.}}{2013}]%
        {cui2013firmware}
\bibfield{author}{\bibinfo{person}{Ang Cui}, \bibinfo{person}{Michael
  Costello}, {and} \bibinfo{person}{Salvatore Stolfo}.}
  \bibinfo{year}{2013}\natexlab{}.
\newblock \showarticletitle{When firmware modifications attack: A case study of
  embedded exploitation}.
\newblock \bibinfo{journal}{\emph{Columbia, Academic Commons}}
  (\bibinfo{year}{2013}).
\newblock


\bibitem[\protect\citeauthoryear{Do, Martini, and Choo}{Do
  et~al\mbox{.}}{2019}]%
        {do2019role}
\bibfield{author}{\bibinfo{person}{Quang Do}, \bibinfo{person}{Ben Martini},
  {and} \bibinfo{person}{Kim-Kwang~Raymond Choo}.}
  \bibinfo{year}{2019}\natexlab{}.
\newblock \showarticletitle{The role of the adversary model in applied security
  research}.
\newblock \bibinfo{journal}{\emph{Computers \& Security}}  \bibinfo{volume}{81}
  (\bibinfo{year}{2019}), \bibinfo{pages}{156--181}.
\newblock


\bibitem[\protect\citeauthoryear{Dodis, Ostrovsky, Reyzin, and Smith}{Dodis
  et~al\mbox{.}}{2008}]%
        {dodis2008fuzzy}
\bibfield{author}{\bibinfo{person}{Yevgeniy Dodis}, \bibinfo{person}{Rafail
  Ostrovsky}, \bibinfo{person}{Leonid Reyzin}, {and} \bibinfo{person}{Adam
  Smith}.} \bibinfo{year}{2008}\natexlab{}.
\newblock \showarticletitle{Fuzzy extractors: How to generate strong keys from
  biometrics and other noisy data}.
\newblock \bibinfo{journal}{\emph{SIAM journal on computing}}
  \bibinfo{volume}{38}, \bibinfo{number}{1} (\bibinfo{year}{2008}),
  \bibinfo{pages}{97--139}.
\newblock


\bibitem[\protect\citeauthoryear{{Dolev} and {Yao}}{{Dolev} and {Yao}}{1983}]%
        {dolevyao}
\bibfield{author}{\bibinfo{person}{D. {Dolev}} {and} \bibinfo{person}{A.
  {Yao}}.} \bibinfo{year}{1983}\natexlab{}.
\newblock \showarticletitle{On the security of public key protocols}.
\newblock \bibinfo{journal}{\emph{IEEE Transactions on Information Theory}}
  \bibinfo{volume}{29}, \bibinfo{number}{2} (\bibinfo{year}{1983}),
  \bibinfo{pages}{198--208}.
\newblock
\urldef\tempurl%
\url{https://doi.org/10.1109/TIT.1983.1056650}
\showDOI{\tempurl}


\bibitem[\protect\citeauthoryear{Eaton}{Eaton}{2017}]%
        {houstonchr}
\bibfield{author}{\bibinfo{person}{Collin Eaton}.}
  \bibinfo{year}{2017}\natexlab{}.
\newblock \bibinfo{title}{Hacked: Energy industry's controls provide an
  alluring target for cyberattacks}.
\newblock
\newblock
\urldef\tempurl%
\url{http://www.houstonchronicle.com/}
\showURL{%
\tempurl}


\bibitem[\protect\citeauthoryear{Falas, Konstantinou, and Michael}{Falas
  et~al\mbox{.}}{2019a}]%
        {falas2019hardware}
\bibfield{author}{\bibinfo{person}{Solon Falas}, \bibinfo{person}{Charalambos
  Konstantinou}, {and} \bibinfo{person}{Maria~K Michael}.}
  \bibinfo{year}{2019}\natexlab{a}.
\newblock \showarticletitle{A Hardware-based Framework for Secure Firmware
  Updates on Embedded Systems}. In \bibinfo{booktitle}{\emph{2019 IFIP/IEEE
  27th International Conference on Very Large Scale Integration (VLSI-SoC)}}.
  IEEE, \bibinfo{pages}{198--203}.
\newblock


\bibitem[\protect\citeauthoryear{Falas, Konstantinou, and Michael}{Falas
  et~al\mbox{.}}{2019b}]%
        {falas2019hardwarespringer}
\bibfield{author}{\bibinfo{person}{Solon Falas}, \bibinfo{person}{Charalambos
  Konstantinou}, {and} \bibinfo{person}{Maria~K Michael}.}
  \bibinfo{year}{2019}\natexlab{b}.
\newblock \showarticletitle{Hardware-Enabled Secure Firmware Updates in
  Embedded Systems}. In \bibinfo{booktitle}{\emph{IFIP/IEEE International
  Conference on Very Large Scale Integration-System on a Chip}}. Springer,
  \bibinfo{pages}{165--185}.
\newblock


\bibitem[\protect\citeauthoryear{Joshi, Mohanty, and Kougianos}{Joshi
  et~al\mbox{.}}{2017}]%
        {joshi2017everything}
\bibfield{author}{\bibinfo{person}{Shital Joshi}, \bibinfo{person}{Saraju~P
  Mohanty}, {and} \bibinfo{person}{Elias Kougianos}.}
  \bibinfo{year}{2017}\natexlab{}.
\newblock \showarticletitle{Everything you wanted to know about PUFs}.
\newblock \bibinfo{journal}{\emph{IEEE Potentials}} \bibinfo{volume}{36},
  \bibinfo{number}{6} (\bibinfo{year}{2017}), \bibinfo{pages}{38--46}.
\newblock


\bibitem[\protect\citeauthoryear{Karri, Sinanoglu, and Rajendran}{Karri
  et~al\mbox{.}}{2017}]%
        {karri2017physical}
\bibfield{author}{\bibinfo{person}{Ramesh Karri}, \bibinfo{person}{Ozgur
  Sinanoglu}, {and} \bibinfo{person}{Jeyavijayan Rajendran}.}
  \bibinfo{year}{2017}\natexlab{}.
\newblock \showarticletitle{Physical unclonable functions and intellectual
  property protection techniques}.
\newblock In \bibinfo{booktitle}{\emph{Fundamentals of IP and SoC Security}}.
  \bibinfo{publisher}{Springer}, \bibinfo{pages}{199--222}.
\newblock


\bibitem[\protect\citeauthoryear{Khan, McLaughlin, Laverty, and Sezer}{Khan
  et~al\mbox{.}}{2017}]%
        {khan2017stride}
\bibfield{author}{\bibinfo{person}{Rafiullah Khan}, \bibinfo{person}{Kieran
  McLaughlin}, \bibinfo{person}{David Laverty}, {and} \bibinfo{person}{Sakir
  Sezer}.} \bibinfo{year}{2017}\natexlab{}.
\newblock \showarticletitle{STRIDE-based threat modeling for cyber-physical
  systems}. In \bibinfo{booktitle}{\emph{2017 IEEE PES Innovative Smart Grid
  Technologies Conference Europe (ISGT-Europe)}}. IEEE, \bibinfo{pages}{1--6}.
\newblock


\bibitem[\protect\citeauthoryear{Kohnfelder and Garg}{Kohnfelder and
  Garg}{1999}]%
        {kohnfelder1999threats}
\bibfield{author}{\bibinfo{person}{Loren Kohnfelder} {and}
  \bibinfo{person}{Praerit Garg}.} \bibinfo{year}{1999}\natexlab{}.
\newblock \showarticletitle{The threats to our products}.
\newblock \bibinfo{journal}{\emph{Microsoft Interface, Microsoft Corporation}}
  \bibinfo{volume}{33} (\bibinfo{year}{1999}).
\newblock


\bibitem[\protect\citeauthoryear{Konstantinou, Keliris, and
  Maniatakos}{Konstantinou et~al\mbox{.}}{2016}]%
        {firmtaxon}
\bibfield{author}{\bibinfo{person}{Charalambos Konstantinou},
  \bibinfo{person}{Anastasis Keliris}, {and} \bibinfo{person}{Michail
  Maniatakos}.} \bibinfo{year}{2016}\natexlab{}.
\newblock \showarticletitle{Taxonomy of firmware Trojans in smart grid
  devices}. In \bibinfo{booktitle}{\emph{Power and Energy Society General
  Meeting (PESGM), 2016}}. IEEE, \bibinfo{pages}{1--5}.
\newblock


\bibitem[\protect\citeauthoryear{Konstantinou and Maniatakos}{Konstantinou and
  Maniatakos}{2015}]%
        {konstantinou2015impact}
\bibfield{author}{\bibinfo{person}{Charalambos Konstantinou} {and}
  \bibinfo{person}{Michail Maniatakos}.} \bibinfo{year}{2015}\natexlab{}.
\newblock \showarticletitle{Impact of firmware modification attacks on power
  systems field devices}. In \bibinfo{booktitle}{\emph{Smart Grid
  Communications, 2015 IEEE International Conference on}}. IEEE,
  \bibinfo{pages}{283--288}.
\newblock


\bibitem[\protect\citeauthoryear{Koushanfar, Boufounos, and Shamsi}{Koushanfar
  et~al\mbox{.}}{2008}]%
        {koushanfar2008post}
\bibfield{author}{\bibinfo{person}{Farinaz Koushanfar}, \bibinfo{person}{Petros
  Boufounos}, {and} \bibinfo{person}{Davood Shamsi}.}
  \bibinfo{year}{2008}\natexlab{}.
\newblock \showarticletitle{Post-silicon timing characterization by compressed
  sensing}. In \bibinfo{booktitle}{\emph{Proceedings of the 2008 IEEE/ACM
  International Conference on Computer-Aided Design}}. IEEE Press,
  \bibinfo{pages}{185--189}.
\newblock


\bibitem[\protect\citeauthoryear{Lab}{Lab}{2019}]%
        {citl2019}
\bibfield{author}{\bibinfo{person}{Cyber Independent~Testing Lab}.}
  \bibinfo{year}{2019}\natexlab{}.
\newblock \bibinfo{title}{Binary Hardening in IoT products}.
\newblock
\newblock
\urldef\tempurl%
\url{https://cyber-itl.org/2019/08/26/iot-data-writeup.html}
\showURL{%
\tempurl}


\bibitem[\protect\citeauthoryear{Mocker}{Mocker}{2019}]%
        {mocker2019}
\bibfield{author}{\bibinfo{person}{Andrijan Mocker}.}
  \bibinfo{year}{2019}\natexlab{}.
\newblock \bibinfo{title}{Tuya: Revised update process hacked again}.
\newblock
\newblock
\urldef\tempurl%
\url{https://www.heise.de/}
\showURL{%
\tempurl}


\bibitem[\protect\citeauthoryear{Moran, Meriac, Tschofenig, and Brown}{Moran
  et~al\mbox{.}}{2019}]%
        {moran2019firmware}
\bibfield{author}{\bibinfo{person}{Brendan Moran}, \bibinfo{person}{Milosch
  Meriac}, \bibinfo{person}{Hannes Tschofenig}, {and} \bibinfo{person}{David
  Brown}.} \bibinfo{year}{2019}\natexlab{}.
\newblock \showarticletitle{A firmware update architecture for Internet of
  Things devices}.
\newblock \bibinfo{journal}{\emph{Internet-Draft
  draft-moran-suit-architecture-02, IETF}} (\bibinfo{year}{2019}).
\newblock


\bibitem[\protect\citeauthoryear{Popp}{Popp}{2009}]%
        {popp2009introduction}
\bibfield{author}{\bibinfo{person}{Thomas Popp}.}
  \bibinfo{year}{2009}\natexlab{}.
\newblock \showarticletitle{An introduction to implementation attacks and
  countermeasures}. In \bibinfo{booktitle}{\emph{2009 7th IEEE/ACM
  International Conference on Formal Methods and Models for Co-Design}}. IEEE,
  \bibinfo{pages}{108--115}.
\newblock


\bibitem[\protect\citeauthoryear{Potkonjak and Goudar}{Potkonjak and
  Goudar}{2014}]%
        {potkonjak2014public}
\bibfield{author}{\bibinfo{person}{Miodrag Potkonjak} {and}
  \bibinfo{person}{Vishwa Goudar}.} \bibinfo{year}{2014}\natexlab{}.
\newblock \showarticletitle{Public physical unclonable functions}.
\newblock \bibinfo{journal}{\emph{Proc. IEEE}} \bibinfo{volume}{102},
  \bibinfo{number}{8} (\bibinfo{year}{2014}), \bibinfo{pages}{1142--1156}.
\newblock


\bibitem[\protect\citeauthoryear{Potkonjak, Meguerdichian, Nahapetian, and
  Wei}{Potkonjak et~al\mbox{.}}{2011}]%
        {potkonjak2011differential}
\bibfield{author}{\bibinfo{person}{Miodrag Potkonjak}, \bibinfo{person}{Saro
  Meguerdichian}, \bibinfo{person}{Ani Nahapetian}, {and}
  \bibinfo{person}{Sheng Wei}.} \bibinfo{year}{2011}\natexlab{}.
\newblock \showarticletitle{Differential public physically unclonable
  functions: architecture and applications}. In
  \bibinfo{booktitle}{\emph{Proceedings of the 48th Design Automation
  Conference}}. \bibinfo{pages}{242--247}.
\newblock


\bibitem[\protect\citeauthoryear{Seals}{Seals}{2019}]%
        {sealstara}
\bibfield{author}{\bibinfo{person}{Tara Seals}.}
  \bibinfo{year}{2019}\natexlab{}.
\newblock \bibinfo{title}{Mirai Botnet Sees Big 2019 Growth, Shifts Focus to
  Enterprises}.
\newblock
\newblock
\urldef\tempurl%
\url{https://threatpost.com/}
\showURL{%
\tempurl}


\bibitem[\protect\citeauthoryear{Soltan, Mittal, and Poor}{Soltan
  et~al\mbox{.}}{2018}]%
        {soltan2018blackiot}
\bibfield{author}{\bibinfo{person}{Saleh Soltan}, \bibinfo{person}{Prateek
  Mittal}, {and} \bibinfo{person}{H~Vincent Poor}.}
  \bibinfo{year}{2018}\natexlab{}.
\newblock \showarticletitle{BlackIoT: IoT Botnet of high wattage devices can
  disrupt the power grid}. In \bibinfo{booktitle}{\emph{27th $\{$USENIX$\}$
  Security Symposium ($\{$USENIX$\}$ Security 18)}}. \bibinfo{pages}{15--32}.
\newblock


\bibitem[\protect\citeauthoryear{Thompson and Zatko}{Thompson and
  Zatko}{2018}]%
        {thompsonzatko2018}
\bibfield{author}{\bibinfo{person}{Parker Thompson} {and}
  \bibinfo{person}{Sarah Zatko}.} \bibinfo{year}{2018}\natexlab{}.
\newblock \showarticletitle{Build Safety of Software in 28 Popular Home
  Routers}.
\newblock \bibinfo{journal}{\emph{Cyber-ITL}} (\bibinfo{date}{Dec}
  \bibinfo{year}{2018}).
\newblock


\bibitem[\protect\citeauthoryear{Wallrabenstein}{Wallrabenstein}{2016}]%
        {wallrabenstein2016practical}
\bibfield{author}{\bibinfo{person}{John~Ross Wallrabenstein}.}
  \bibinfo{year}{2016}\natexlab{}.
\newblock \showarticletitle{Practical and secure IoT device authentication
  using physical unclonable functions}. In \bibinfo{booktitle}{\emph{2016 IEEE
  4th International Conference on Future Internet of Things and Cloud
  (FiCloud)}}. IEEE, \bibinfo{pages}{99--106}.
\newblock


\bibitem[\protect\citeauthoryear{{Yanambaka}, {Mohanty}, {Kougianos}, and
  {Puthal}}{{Yanambaka} et~al\mbox{.}}{2019}]%
        {puf2019iomt}
\bibfield{author}{\bibinfo{person}{V.~P. {Yanambaka}}, \bibinfo{person}{S.~P.
  {Mohanty}}, \bibinfo{person}{E. {Kougianos}}, {and} \bibinfo{person}{D.
  {Puthal}}.} \bibinfo{year}{2019}\natexlab{}.
\newblock \showarticletitle{PMsec: Physical Unclonable Function-Based Robust
  and Lightweight Authentication in the Internet of Medical Things}.
\newblock \bibinfo{journal}{\emph{IEEE Transactions on Consumer Electronics}}
  \bibinfo{volume}{65}, \bibinfo{number}{3} (\bibinfo{year}{2019}),
  \bibinfo{pages}{388--397}.
\newblock


\end{thebibliography}

\end{document}